\documentclass[preprint,aps,12pt,showpacs,nofootinbib,tightenlines]{revtex4}
\usepackage{amsmath}
\usepackage{amssymb}
\usepackage{epsfig}
\usepackage{graphicx}
\begin{document}
\preprint{ITP-TH-2003-17}
\newcommand{\beq}{\begin{eqnarray}}
\newcommand{\eeq}{\end{eqnarray}}

\newcommand{\bxsga}{B\to X_s \gamma}
\newcommand{\brbxsga}{{\cal B}(B\to X_s \gamma)}
\newcommand{\bzbzb}{ B_d^0 - \bar{B}_d^0 }

\newcommand{\bsga}{  b\to s \gamma}
\newcommand{\bdga}{  b\to d \gamma}
\newcommand{\bvga}{  B\to V \gamma }
\newcommand{\bksga}{ B\to K^* \gamma}
\newcommand{\brhoga}{B\to \rho \gamma}

\newcommand{\brbkz}{{\cal B}(B\to \overline{K}^{*0} \gamma)}
\newcommand{\brbkm}{{\cal B}(B\to K^{*-} \gamma)}
\newcommand{\brbrm}{{\cal B}(B\to \rho^- \gamma)}
\newcommand{\brbrz}{{\cal B}(B\to \rho^0 \gamma)}

\newcommand{\calb}{ {\cal B}}
\newcommand{\acp}{ {\cal A}_{CP}}
\newcommand{\oas}{ {\cal O} (\alpha_s)}

\newcommand{\mt}{m_t}
\newcommand{\mw}{M_W}
\newcommand{\mhp}{M_{H}}
\newcommand{\muw}{\mu_W}
\newcommand{\mub}{\mu_b}
\newcommand{\dmd}{\Delta M_{B_d} }
\newcommand{\ltt}{\lambda_{tt} }
\newcommand{\lbb}{\lambda_{bb} }
\newcommand{\rhob}{\bar{\rho} }
\newcommand{\etab}{\bar{\eta} }

\newcommand{\smallsm}{{\scriptscriptstyle SM}}
\newcommand{\smallyy}{{\scriptscriptstyle YY}}
\newcommand{\smallxy}{{\scriptscriptstyle XY}}
\newcommand{\smallnp}{{\scriptscriptstyle NP}}

\newcommand{\tab}[1]{Table \ref{#1}}
\newcommand{\fig}[1]{Fig.\ref{#1}}
\newcommand{\real}{{\rm Re}\,}
\newcommand{\im}{{\rm Im}\,}
\newcommand{\non}{\nonumber\\ }

\def \epjc{  Eur. Phys. J. C }
\def \jpg{  J. Phys. G}
\def \npb{  Nucl. Phys. B }
\def \plb{  Phys. Lett. B }
\def \prd{  Phys. Rev. D }
\def \prl{  Phys. Rev. Lett.  }
\def \pr{   Phys. Rep. }
\def \rmp{  Rev. Mod. Phys. }
\title{Exclusive $B \to PV $ Decays and CP Violation in the General two-Higgs-doublet Model}
\author{ Y.L.Wu}
\email{ylwu@itp.ac.cn}
\author{C. Zhuang }
\email{zhuangc@itp.ac.cn} \affiliation{ Kavli Institute for
Theoretical Physics China (KITPC) \\ Institute of theoretical
physics,Chinese Academy of Science, Beijing,100080, P.R.China}
%
\date{\today}
\begin{abstract}
We calculate all the branching ratios and direct CP violations of $B
\to PV$ decays in a most general two-Higgs-doublet model with
spontaneous CP violation. As the model has rich CP-violating
sources, it is shown that the new physics effects to direct CP
violations and branching ratios in some channels can be significant
when adopting the generalized factorization approach to evaluate the
hadronic matrix elements, which provides good signals for probing
new physics beyond the SM in the future B experiments.

\end{abstract}

\pacs{12.60.Fr;13.25.Hw;11.30.Hv;}

\maketitle

\newpage

\section{introduction}

    To understand the origin of CP violation (CPV) is an important
    subject not only for exploring the basic symmetry of space-time
    and elementary particles but also for understanding the evolution of our
    universe. It is well known that in the Standard Model(SM) of particle physics, CP
    violation is characterized by a single weak phase in Cabibbo-Kobayashi-Maskawa
    matrix\cite{KM}, which can provide a well explanation for the direct CP violation
    $\varepsilon'/\varepsilon$ \cite{YLWU} established in kaon decays\cite{KK},
    and also direct CP violation\cite{WZZ} observed in B-meson decays\cite{BB}.
    Though the theory of the strong and electroweak(EW) interactions
    in SM has met with extraordinary success, it is widely believed that
    the SM can not be the final theory of particle physics, in particular
    because the Higgs sector of SM is not well understood yet and the CP phase in CKM matrix
    is not enough to understand the baryon and anti-baryon asymmetry
    in the universe. It was suggested that CP symmetry may be broken down spontaneously\cite{TDL}.
    Many possible extensions of SM in Higgs sector have been
    proposed\cite{HS}.
    Other possible extensions of the SM have been explored,
    such as the Super Symmetric model(SUSY), little Higgs model
    and extra dimensions, which all make better the situation of the SM.
    But no single model is good enough to solve all the
    problems existing in the SM and it is then worthwhile to consider
    all the possibilities beyond SM. As one of the simplest extensions of the SM, the so-called
   two-higgs-doublet model(2HDM) which introduces an extra Higgs doublet without imposing the {\it ad hoc}
   discrete symmetries has been investigated widely from various considerations\cite{TH1,TH2,TH3,TH4,YLW1,YLW2,ARS,dai,BCK,KSW}.
   Motivated solely from the origin of CP violation, a general two Higgs doublet model with
   spontaneous CP violation (Model III 2HDM) has been shown to provide one of
   the simplest and attractive models in understanding the origin and
    mechanism of CP violation at weak scale\cite{YLW1,YLW2}. In such a model,
    there exists more physical neutral and charged Higgs bosons and
    rich induced CP-violating sources from a single CP phase of vacuum. Of particular, the model III 2HDM allows
    flavor-changing neutral currents but suppressed by approximate U(1) flavor symmetry,
    which is different from the so-called model I and model II 2HDM in which a \textit{ad-hoc} discrete
    symmetry ($Z_2$ symmetry) has been imposed to avoid the FCNC.

    It is known that the FCNC's concerning
    the first two generations are highly suppressed from
    low-energy experiments, and those involving the third generation is not as severely
    suppressed as the first two generations. So the model III 2HDM can
    be parameterized in a way to satisfy the current experimental constraints.
    The constraints on Model III 2HDM from neutral meson mixings ($K^0-\bar{K}^0$, $D^0-\bar{D}^0$,
    $B^0-\bar{B}^0$)\cite{WZ} and radiative decays of bottom quark\cite{WW,YLW3,hcs3} have been studied in
    details. In this note, we shall investigate the possible new effects of model III 2HDM on
    two-body charmless nonleptonic B decays $B \to h_1h_2$ with
    $h_1, h_2$ being the charmless light hadrons.
    This is because those decays have triggered considerable theoretical interest in
    understanding SM. Also those decay channels are also thought to be sensitive and important in exploring new physics
    beyond the SM as they involve the so-called tree (current-current) $b\to (u,c)$
    and/or $B\to (d,s)$ penguin amplitudes with both QCD and electroweak penguin transition participating.
    In the 2HDM, there are five Higgs particles including
    the $H^0$ Higgs in SM, these extra Higgs will mediate all the
    penguin transitions. As the couplings involving Higgs
    bosons and fermions have complex CP phases in the model III 2HDM,
    CP violation effects occur even in the simplest case that all
    the tree-level FCNC couplings are negligible. With the improvement of
    experimental precision, more and more direct CPV have been observed and will
    be much precisely tested in the future experiments.

    The paper is organized as follows. In Sec II, we first describe the
    theoretical frame including a brief introduction of the
    two-Higgs-doublet model with spontaneous CP violation, i.e., Model III 2HDM,
    and the effective Hamiltonian as well as the
    generalized factorization formula, which is our basic tool to
    estimate the branching ratios and CPV asymmetry of B meson
    decays. In Sec III, we make a detailed calculation with numerical results evaluated from
    a factorization ansatz which allows us to express the matrix
    elements $<h_1h_2|H_{eff}|B>$ as a product of two factors
    $<h_1|J_1|B><h_2|J_2|B>$, and make quantitative predictions.
    Our conclusions and discussions are presented in the last section.

\section{Theoretical Framework}

\subsection{Outline of the Two-Higgs-doublet Model}

  One of the important developments of SM is the so-called Higgs mechanism,
  i.e., a spontaneous symmetry breaking mechanism by which the gauge bosons and
  fermions can get their masses. In the SM, a single Higgs doublet of SU(2) is sufficient
  to break the  $SU(2)_L\times U(1)_Y$ symmetry to $U(1)_{em}$ and generate mass to the gauge bosons and fermions.
  Nevertheless, the Higgs sector of the SM has not been experimentally tested although enormous
  efforts have been made. For the origin of CP violation, SM gives no
  explanation as there is only one single neutral Higgs in SM
  and its interaction coupling constants are fixed by the known
  parameters and the fermion masses. Many attempts have been made by both theoretists and experimentalists
  to explore the mechanisms of CP violation since the discovery of CP violation in 1964.
  Spontaneous CP violation requires at least two Higgs doublets. A consistent and simple model which
  provides a spontaneous CP violation mechanism was constructed completely
  in a general Two-Higgs-doublet model\cite{YLW1,YLW2}. Such a model III 2HDM not only explains the origin
  of CP violation in the SM, but also induces rich new resources of
  CP violation. The new sources of CP violation can lead to some new phenomenological effects
  which are promising to be tested by the future B factory and LHCb. In this note,
  we will focus on the phenomenological applications of the model III 2HDM
  in the two-body charmless hadronic $B\to PV$ decays.

  The two complex Higgs doublets in the Model III 2HDM are expressed
  as\cite{YLW1,YLW2,ARS,BCK,KSW}:
  \beq
 \phi_1 =  \left(\begin{array}{l}
               \phi_1^\dag \\
               \phi_1^0 \\
               \end{array}\right),
 \phi_2 =  \left(\begin{array}{l}
               \phi_2^\dag \\
               \phi_2^0 \\
               \end{array}\right)
 \eeq
 The corresponding Higgs potential can simply be written in the following
 general form:
 \beq
 V(\phi) &=& \lambda_1(\phi_1^\dag\phi_1-\frac{1}{2}v_1^2)^2 +
                 \lambda_2(\phi_2^\dag\phi_2-\frac{1}{2}v_2^2)^2\non
                 &&+\lambda_3(\phi_1^\dag\phi_1-\frac{1}{2}v_1^2)(\phi_2^\dag\phi_2-\frac{1}{2}v_2^2)+\lambda_4[(\phi_1^\dag\phi_2)(\phi_2^\dag\phi_1)]\non
                 &&+\frac{1}{2}\lambda_5(\phi_1^\dag\phi_2+\phi_2^\dag\phi_1-v_1v_2\cos{\delta})^2+\lambda_6(\phi_1^\dag\phi_2-\phi_2^\dag\phi_1-v_1v_2\sin{\delta})^2\non
                 &&+[\lambda_7(\phi_1^\dag\phi_1-\frac{1}{2}v_1^2)^2+\lambda_8(\phi_2^\dag\phi_2-\frac{1}{2}v_2^2)^2][\phi_1^\dag\phi_2+\phi_2^\dag\phi_1-v_1v_2\cos{\delta}],
 \eeq
 where $\lambda_i (i = 1,2,...8)$ are all real parameters. If
 all $\lambda_i$ are non-negative, the minimum occurs at :
 \beq
 <\phi_1^0> = v_1e^{i\delta},\non
 <\phi_2^0> = v_2,
 \eeq
   With $v_1, v_2$ are the vacuum expectation values of $\phi_1, \phi_2$ respectively, and
   $\delta$ the relative phase of the vacuum. It is clear that in the above potential CP nonconservation can
    only occur through the vacuum with $\delta \neq 0$.
    Obviously, such a CP violation appears as an explicit one in the
    potential when $\lambda_6 \neq 0$\cite{YLW2}.

    After a unitary transformation, it is natural and convenient
     to use the following basis:
     \beq
   H_1 &=&  \frac{1}{\sqrt{2}}\left[\begin{array}{l}
               \sqrt{2}G^+ \\
               v+\phi_1^0+iG^0 \\
               \end{array}\right],\non
 H_2 &=&  \frac{1}{\sqrt{2}}\left[\begin{array}{l}
               \sqrt{2}H^+ \\
               \phi_2^0+iA^0 \\
               \end{array}\right],
 \eeq
  with:
 \beq
 <H_1^0> &=& ve^{i\delta},\non
 <H_2^0> &=& 0,
 \eeq
 where $v = \sqrt{v_1^2+v_2^2}$ and is related to the $W$ mass by
 $M_W = gv/2$. Here $H^0$ plays the role of the Higgs boson in the standard
 model. $H^\pm$ are the charged scalar pair with $H^\pm
 = \sin{\beta}\phi_1^\pm e^{-i\delta} - \cos{\beta}\phi_2^\pm$,
 where $\tan{\beta} = v_2/v_1$. And as for the neutral Higgs,
 $\phi_1, \phi_2$ are not the neutral mass eigenstates but linear
 combinations of CP-even neutral Higgs boson mass eigenstates,
 $H_0$ and $h_0$:
 \beq
 H_0 &=& \phi_1^0\cos{\alpha}+\phi_2^0\sin{\alpha},\non
 h_0 &=& -\phi_1^0\sin{\alpha}+\phi_2^0\cos{\alpha},
 \eeq
 where  $\alpha$ is the mixing angle and when $\alpha = 0$, $(\phi_1^0,
 \phi_2^0)$ are identical with $(H_0, h_0)$. For simplicity, the
 mixing with the pseudoscalar $A^0$ is not considered here.

 Let us consider a Yukawa Lagrangian of the following form:
 \beq
{\cal L}_Y &=& \xi^U_{1ij}\bar{Q}_{i,L} \tilde{\phi_1}U_{j,R} +
\xi^D_{1ij}\bar{Q}_{i,L} \phi_1 D_{j,R} +\xi^U_{2ij}\bar{Q}_{i,L}
\tilde{\phi_2}U_{j,R} +\xi^D_{2ij}\bar{Q}_{i,L} \phi_2 D_{j,R}+
H.c., \label{leff} \eeq where $\phi_i (i = 1, 2)$ are the two Higgs
doublets, .$\widetilde{\phi}_{1,2}= i\tau_2\phi_{1,2}^*$, $Q_{i,L}
(U_{j,R})$ with $i = (1, 2, 3)$ are the left-handed isodoublet
quarks (right-handed up-type quarks), $D_{j,R}$ are the righthanded
isosinglet down-type quarks, while $\xi^{U,D}_{1ij}$ and
$\xi^{U,D}_{2ij}$ ($i, j = 1, 2, 3$ are family index ) are generally
the nondiagonal matrices of the Yukawa coupling. After diagonalizing
the mass matrix of quark fields, the Yukawa Lagrangian that related
to the decays we considered in this paper can be written as:
 \beq {\cal L}_Y &=&
-\frac{g}{2M_W}(H^0\cos\alpha-h^0\sin\alpha)(\overline{U}M_UU+\overline{D}M_DD)\non
 &&
 -\frac{H^{0}\sin\alpha+h^0\cos\alpha}{\sqrt{2}}[\overline{U}(\xi^U\frac{1}{2}
 (1+\gamma_5)+\xi^{U\dag}\frac{1}{2}(1-\gamma_5))U\non
 &&+\overline{D}(\xi^D\frac{1}{2}(1+\gamma_5)+\xi^{D\dag}\frac{1}{2}(1-\gamma_5))D]\non
 &&+
 \frac{iA^0}{\sqrt{2}}[\overline{U}(\xi^U\frac{1}{2}(1+\gamma_5)
 -\xi^{U\dag}\frac{1}{2}(1-\gamma_5))U-\overline{D}(\xi^D\frac{1}{2}
 (1+\gamma_5)-\xi^{D\dag}\frac{1}{2}(1-\gamma_5))D]\non
 &&-H^+\overline{U}[V_{CKM}\xi^D\frac{1}{2}(1+\gamma_5)-\xi^{D\dag}\frac{1}{2}(1-\gamma_5)]D\non
 &&-H^-\overline{D}[\xi^{D\dag}V_{CKM}^\dag\frac{1}{2}(1-\gamma_5)-V_{CKM}^\dag \xi^{U}\frac{1}{2}(1+\gamma_5)]U,
 \eeq

 where $U$ represents the mass eigenstates of $u,c,t$ quarks and $D$ represents the mass eigenstates of $d,s,b$
 quarks, $V_{CKM}$ is the Cabibbo-Kobayashi-Maskawa matrix and
 $\hat{\xi}^{U,D}$ are the FCNC couplings in the mass eigenstate, and they may be
 parameterized in terms of the quark mass:
 \beq
  \xi_{ij}^{U,D} &=&
  \lambda_{ij}\frac{g\sqrt{m_im_j}}{\sqrt{2}M_W},\non
  \hat{\xi}^U &=& \xi^{U} \cdot V_{CKM},\non
  \hat{\xi}^D &=& V_{CKM} \cdot \xi^{D} ,
  \eeq
  The first two generations' FCNC are naturally suppressed by
  the small quark masses,but the third generation has more space to get FCNC contributions.
  In this paper, we just choose the $\xi^{U,D}$ to be
  diagonal $\xi_{ii}^{U,D} \equiv \xi_i^{U,D}$ $(i=s,c,b,t)$ , and neglect the first generation quarks'
  contributions.So the really leading contribution arises from the
  diagram with a top quark in the loop and the relevant couplings
  will be $\hat{\xi}_{ts}^{U,D}$ and $\hat{\xi}_{tb}^{U,D}$, they are explicitly given by:
  \beq
 \hat{\xi}_{ts}^U &=& \xi_{t}^U V_{ts},\quad   \hat{\xi}_{tb}^U = \xi_{t}^U
 V_{tb}\non
 \hat{\xi}_{ts}^D &=& \xi_{s}^D V_{ts},\quad   \hat{\xi}_{tb}^D = \xi_{b}^D V_{tb},
 \eeq
  From the above parameterization, the free parameters in this model are $\lambda_{ij}(i,j = s,c,t,b)$.
  Their values can be constrained through experiments.

  In the model III 2HDM with spontaneous CP violation, the induced CP violation can
  be classified into the following four types via their interactions\cite{YLW1,YLW2}: i) from the CKM matrix;
  ii) from the charged Higgs couplings to the fermions $\xi_{charged}$;
  iii) from the neutral Higgs couplings to the fermions $\xi_{neutral}$;
   iv) from the CP nonconservation Higgs potential
    $V(\phi)$ via mixings among scalars and pseudoscalar bosons.

  The model allows flavor-changing-neutral-currents(FCNC) at tree
  level and via loop effects due to exchanges of Higgs bosons.
  One of the most stringent tests is from the
  radiative decay of B mesons and also from the inclusive decay
  rate of $b \to s\gamma$ which has the least hadronic
  uncertainties. Other constraints could come from the $B_0-\overline{B_0}$ mixing, $\rho_0$, $R_b$ and the
  neutron electric dipole moment etc. In this note, we shall consider possible new effects
  in charmless hadronic two body decays of bottom mesons.

 \subsection{Effective Hamiltonian and Wilson coefficients}

  The effective Hamiltonian for charmless B decays with $\Delta B =
  1$ is:
  \beq
  {\cal H}
  &=&\frac{G_F}{\sqrt{2}}\sum_{p=u,c}V_{pb}V_{ps}^*(C_1Q_1^p+C_2Q_2^p+\sum_{i=3,...,16}[C_iQ_I+C_i^{'}Q_i^{'}]\non
  && +C_{7\gamma}Q_{7\gamma}+C_{8g}Q_{8g}+
  C_{7\gamma}^{'}Q_{7\gamma}^{'}+C_{8g}^{'}Q_{8g}^{'})+h.c.\label{heff}
  \eeq

The operators $Q_{1,...10},Q_{7\gamma},Q_{8g}$ can be found in
\cite{buchalla1}, of which the $Q_1$ and $Q_2$ are the
current-current operators and  $Q_3 - Q_6$ are QCD penguin
operators. $Q_{7\gamma}$ and $Q_{8g}$ are, respectively, the
magnetic penguin operators for $b \to s\gamma$ and $b \to sg$. Here
the mass of the external strange quark is neglected compared to the
external bottom-quark mass.


 The additional new operators related to the neutral
Higgs mediated processes($b \to s\overline{q}q $ )are\cite{hcs1}:
\beq
 Q_{11} &=& (\overline{s}b)_{(S+P)}\sum(\overline{q}q)_{(S-P)},\non
 Q_{12} &=& (\overline{s}_ib_j)_{(S+P)}\sum(\overline{q}_jq_i)_{(S-P)},\non
 Q_{13} &=& (\overline{s}b)_{(S+P)}\sum(\overline{q}q)_{(S+P)},\non
 Q_{14} &=& (\overline{s}_ib_j)_{(S+P)}\sum(\overline{q}_jq_i)_{(S+P)},\non
 Q_{15} &=& (\overline{s}\sigma^{\mu\nu}(1+\gamma_5)b)\sum(\overline{q}\sigma^{\mu\nu}(1+\gamma_5)q),\non
 Q_{16} &=& (\overline{s}_i\sigma^{\mu\nu}(1+\gamma_5)b_j)\sum(\overline{q}_j\sigma^{\mu\nu}(1+\gamma_5)q_i),\non
 \eeq
 where $(\bar{q_1}q_2)_{S\pm P} = \bar{q_1}(1\pm \gamma_5)q_2$, $q = u,d,s,c,b$.
 The operators $Q_i^{'}$ in Eq(\ref{heff}) are obtained from the $Q_i$ by exchanging
 $L\leftrightarrow R$. As the primed operators's contributions
 are suppressed by $m_s/m_b$, we shall neglect their effects in our present considerations.
 The Wilson Coefficients $C_i, i=1,...10$ have been calculated at LO\cite{LO1,LO2} and NLO
 \cite{buchalla1} in SM and also at LO in 2HDM\cite{wise,xiao1}. Here we list their initial
coefficient functions
 in the 2HDM\cite{zhuang,xiao1}:
 \beq
 C_1(M_W) &=& \frac{11}{2}\frac{\alpha_s(M_W)}{4\pi},\non
 C_2(M_W) &=&
 1-\frac{11}{6}\frac{\alpha_s(M_W)}{4\pi}-\frac{35}{18}\frac{\alpha}{4\pi},\non
 C_3(M_W) &=&
 -\frac{\alpha_s(M_W)}{24\pi}\{\widetilde{E}_0(x_t)+E_0^{III}(y)\}+\frac{\alpha}{6\pi}[2B_0(x_t)+C_0(x_t)],\non
 C_4(M_W) &=&
 \frac{\alpha_s(M_W)}{8\pi}\{\widetilde{E}_0(x_t)+E_0^{III}(y)\},\non
 C_5(M_W) &=&
 -\frac{\alpha_s(M_W)}{24\pi}\{\widetilde{E}_0(x_t)+E_0^{III}(y)\},\non
 C_6(M_W) &=&
 \frac{\alpha_s(M_W)}{8\pi}\{\widetilde{E}_0(x_t)+E_0^{III}(y)\},\non
 C_7(M_W) &=&
 \frac{\alpha(M_W)}{6\pi}[4C_0(x_t)+\widetilde{D}_0(x_t)],\non
  C_8(M_W) &=& 0,\non
  C_9(M_W) &=&
  \frac{\alpha}{6\pi}[4C_0(x_t)+\widetilde{D}_0(x_t)+\frac{1}{\sin^2\theta_W}(10B_0(x_t)-4C_0(x_t))],\non
 C_{10}(M_W) &=& 0,
 \eeq
 and the LO $C_{7\gamma}, C_{8g}$ are sufficient:
 \beq
 C_{7\gamma}(M_W) &=&
 -\frac{A(x_t)}{2}-\frac{A(y)}{6}|\lambda_{tt}|^2+B(y)|\lambda_{tt}\lambda_{bb}|e^{i\theta},\non
 C_{8g}(M_W) &=& -\frac{D(x_t)}{2}-\frac{D(y)}{6}|\lambda_{tt}|^2+E(y)|\lambda_{tt}\lambda_{bb}|e^{i\theta},
 \eeq
  where $x_t = m_t^2/M_W^2$, and $y = m_t^2/M_{H^{\pm2}}$. The
  Inami-Lim functions $A,B,D,E......$ are known in SM and 2HDM\cite{wise}:

 For the new operators $Q_{(11,12...16)}$, the corresponding
 Wilson coefficients $C_i,i=11,...16$  at leading order have been calculated in\cite{hcs1,hcs2}:
 \beq
 C_{11}(M_W) &=&
 \frac{\alpha}{4\pi}\frac{m_b}{m_{\tau}\lambda_{\tau\tau}^*}(C_{Q_1}-C_{Q_2}),\non
 C_{13}(M_W) &=&
 \frac{\alpha}{4\pi}\frac{m_b}{m_{\tau}\lambda_{\tau\tau}}(C_{Q_1}+C_{Q_2}),\non
 C_{12}(M_W) &=& C_{14}(M_W) = C_{15}(M_W) = C_{16}(M_W) = 0 ,
 \eeq
 Here the explicit expression of $C_{Q_1}, C_{Q_1} $ can be found
 in \cite{hcs2}.

 For the $B \to PV$ processes, the Wilson coefficient functions must
 run from the $M_W$ scale to the scale of $O(m_b)$. For $C_1 - C_{10}$, the NLO corrections should
 be included. While for $C_{8g}$ and $C_{7\gamma}$, LO results are
 sufficient. The details for the running Wilson coefficients
 can be found in Ref.\cite{buchalla1}. As for the neutral Higgs boson induced
 operators, the one loop anomalous dimension matrices can be
 divided into two distangled groups\cite{hcs1}:
 \beq
 \gamma^{RL} &=& \begin{array}{c|cc}
     & O_{11} & O_{12} \\
     \hline
   O_{11} & -16 & 0 \\
   O_{12} & -6 & 2 \\
 \end{array}
 \eeq
 and
\beq
\gamma^{RR} &=&\begin{array}{c|cccc}
     & O_{13} & O_{14} & O_{15} & O_{16} \\
     \hline
   O_{13} & -16 & 0 & 1/3 & -1  \\
   O_{14} & -6 & 2 & -1/2 & -7/6 \\
   O_{15} & 16 & -48 & 16/3 & 0 \\
   O_{16} & -24 & -56 & 6 & -38/3 \\
 \end{array}
\eeq
 As no NLO Wilson coefficients $C_i, i= 11,12,...16$ are
available, we may just use the LO Wilson coefficients for a
numerical estimation.

 \subsection{Generalized Factorization formula}

   For our present purpose, we may use the generalized factorization
   method\cite{hyc1,hyc2,ali,lucd} to evaluate the hadronic matrix elements.
   We know that in full theory, the leading order QCD corrections to the weak
   transition is of the form $\alpha_s\ln(M_W^2/-p^2)$ for massless
   quarks, where $p$ is the off-shell momentum of external quark
   lines and depends on the system under consideration.We can
   choose a renormalization scale $\mu$ and separate $\ln(M_W^2/-p^2) =
   \ln(M_W^2/\mu^2)+\ln(\mu^2/-p^2)$. The first part
   $\ln(M_W^2/\mu^2)$is included in the Wilson coefficients $c(\mu)$and
   summed over to all orders in $\alpha_s$ using the renormalization
   group equation, while the second part is due to the matrix
   element evaluations and is small. It is related to the tree
   matrix element via:
   \beq
   \langle O(\mu) \rangle &=& g(\mu)\langle O \rangle_{tree}
   \eeq
   with:
   \beq
   g(\mu) \sim 1+\alpha_s(\mu)(\gamma\ln\frac{\mu^2}{-p^2}+c)
   \eeq
   where the $\mu$ dependence of the matrix elements is approximately extracted out
   to the function $g(\mu)$, that is:
   \beq
   \langle {\cal H}_{eff} \rangle &=& c(\mu)g(\mu)\langle O
   \rangle_{tree} = c^{eff}\langle O
   \rangle_{tree}
   \eeq
   the effective Wilson coefficients $c^{eff}$ should  be
   in principle renormalization scale independent. Thus it is necessary to
   incorporate QCD and EW corrections to the operators:
   \beq
  \langle O_i(\mu) \rangle &=&
  [I+\frac{\alpha_s(\mu)}{4\pi}\hat{m_s}(\mu)+\frac{\alpha}{4\pi}\hat{m_e}(\mu)]_{ij}\langle O
  \rangle_{tree},
 \eeq
 with
 \beq
 c_i^{eff} &=&
 [I+\frac{\alpha_s(\mu)}{4\pi}\hat{m_s}^T(\mu)+\frac{\alpha}{4\pi}\hat{m_e}^T(\mu)]_{ij}c_j(\mu),
 \eeq
  The perturbative QCD and EW corrections to the matrices
  $\hat{m_s}$ and $\hat{m_e}$ from the vertex and penguin diagrams
  can be found in\cite{ali,du,leutwyler}.

  Using the following parameterization for decay constant and form
  factors:
  \beq
  <0|A_{\mu}|P(q)> &=& if_Pq_{\mu},  <0|V_{\mu}|V(p,\epsilon)> =
  f_Vm_V\epsilon_{\mu},\non
  \eeq
  we arrive at
  \beq
  X^{BP,V} \equiv
  <V|(\bar{q_2}q_3)_{V-A}|><P|(\bar{q_1}b)_{V-A}|B> &=&
  2f_Vm_VF_1^{B \to P}(m_V^2)(\epsilon\cdot p_{B}),\non
 X^{BV,P} \equiv
  <P|(\bar{q_2}q_3)_{V-A}|><V|(\bar{q_1}b)_{V-A}|B> &=&
  2f_Pm_VA_0^{B \to V}(m_P^2)(\epsilon\cdot p_{B}),
  \eeq
  Using the Fierz Transformation,
  \beq
  (V-A)(V+A) \rightarrow -2(S-P)(S+P),\non
  (V-A)(V-A) \rightarrow (V-A)(V-A)
  \eeq
  One can easily obtain all the $Q_{1,...10}$ tree level matrix elements\cite{hyc1,hyc2}.
  For the new operators $Q_{11,...16}$, the additional factorization formulas are\cite{ball2}:
  \beq
  <V(k,\epsilon^*)|\bar{q}\sigma^{\mu\nu}q'|0> &=&
  -i(\epsilon_\mu^*k_\nu-\epsilon_\nu^*k_\mu)f_V^\bot, \non
  <P(p)|\bar{q}\sigma^{\mu\nu}k^\nu q'|B(p_B)> &=&
  \frac{i}{m_B+m_P}\{q^2(p+p_B)_{\mu}-(m_B^2-m_P^2)q_\mu\}f_T^P,
   \eeq
  with $k=p_B-P$ and $q= p_B -p$. $f_V^{\bot}$ and $f_T^P$ are the tensor decay constant of vector meson and
   the tensor form factor relevant to $B \to P$ decays.
   $\epsilon^*$ is the polarization vector of vector meson.
   The hadronic matrix element is given by
  \beq
  <V(k,\epsilon^*)|\overline{q^{'}}\sigma^{\mu\nu}q|0><
  P(p)|\overline{q}\sigma_{\mu\nu}b|B(p_B)> &=&
  \frac{2f_V^\bot f_T^Pm_V^2}{m_B+m_P}(\epsilon^*\cdot p_B),
  \eeq

The tree level matrix elements of $Q_{(11,12,...16)}$ can be
factorized as ($b \to s$ for example):
 \beq
  <P V|Q_{11}|B> &=& a_{11}\frac{m_P^2}{(m_b+m_s)(m_q+m_{q'})}<P|(\overline{q'}q)_{V-A}| 0>< V|(\overline{s}b)_{V-A}|B>, \non
  <P V|Q_{12}|B> &=& -\frac{1}{2}a_{12}<P|(\overline{q'}q)_{V-A}| 0><V|(\overline{s}b)_{V-A}|B> \non
                &=& \frac{1}{2}a_{12}<V|(\overline{q'}q)_{V-A}| 0><P|(\overline{s}b)_{V-A}|B>,\non
  <P V|Q_{13}|B> &=& -a_{13}\frac{m_P^2}{(m_b+m_s)(m_q+m_{q'})}<P|(\overline{q'}q)_{V-A}| 0>< V|(\overline{s}b)_{V-A}|B>, \non
  <P V|Q_{14}|B> &=& -\frac{1}{2}a_{14}\frac{m_P^2}{(m_b+m_s)(m_q+m_{q'})}<P|(\overline{q'}q)_{V-A}|
  0><V|(\overline{s}b)_{V-A}|B>\non
  &=&\frac{1}{4}<|V\overline{q^{'}}\sigma^{\mu\nu}q|0><
  P|\overline{s}\sigma_{\mu\nu}b|B>, \non
  <P V|Q_{15}|B> &=& 2a_{15}<|V\overline{q^{'}}\sigma^{\mu\nu}q|0><
  P|\overline{s}\sigma_{\mu\nu}b|B>, \non
  <P V|Q_{16}|B> &=& -a_{16}<|V\overline{q^{'}}\sigma^{\mu\nu}s|0><
  P|\overline{q}\sigma_{\mu\nu}b|B>\non
   &=&6a_{16}\frac{m_P^2}{(m_b+m_q)(m_{q'}+m_s)}<P|(\overline{q'}s)_{V-A}| 0><V|(\overline{q}b)_{V-A}|B>, \non
  \eeq
with:
   \beq
   a_{11} &=& c_{11}+\frac{c_{12}}{N_c^{'}},\quad a_{12} = c_{12}+\frac{c_{11}}{N_c^{'}},\non
   a_{13} &=& c_{13}+\frac{c_{12}}{N_c^{'}},\quad a_{14} = c_{14}+\frac{c_{13}}{N_c^{'}},\non
   a_{15} &=& c_{15}+\frac{c_{16}}{N_c^{'}},\quad a_{16} = c_{16}+\frac{c_{15}}{N_c^{'}},
   \eeq
$N_c'$ is the effective color number relative to the new six
operators,which is set to be universal in all the decay
channels.In this paper we fix it to be $N_c'=3$ to estimate the
neutron higgs effects.
  As for the SM operators,besides the perturbative QCD and EW corrections to the hadronic
  matrix elements that can be factorized into the effective Wilson coefficients, there still exists the
  nonfactorizable effects, such as the spectator quark effects,
  annihilation diagrams and space-like penguins. Consider an
  arbitrary operator of the form $O = \bar{q_1}^\alpha\Gamma q_2^\beta \bar{q_3}^\beta\Gamma' q_4^\alpha$
  which arises from the Fierz transformation of a singlet-singlet
  operator with $\Gamma$ and $\Gamma'$ being some combinations of
  Dirac matrices. By using the identity:
  \beq
  O = \frac{1}{3}\bar{q_1}\Gamma q_2 \bar{q_3}\Gamma' q_4 + \frac{1}{2}\bar{q_1}\lambda^\alpha\Gamma q_2 \bar{q_3}\lambda^\alpha\Gamma'
  q_4,
  \eeq
  the matrix element of $M \to P_1P_2$ can be expanded as:
  \beq
  <P_1P_2|O|M> &=& \frac{1}{3}<P_1|\bar{q_1}\Gamma q_2|0><P_2|\bar{q_3}\Gamma' q_4
  |M>_f + \frac{1}{3}<P_1|\bar{q_1}\Gamma q_2|0><P_2|\bar{q_3}\Gamma' q_4
  |M>_{nf}\non
  &&+ \frac{1}{2}<P_1P_2|\bar{q_1}\lambda^\alpha\Gamma q_2 \bar{q_3}\lambda^\alpha\Gamma'
  q_4|M>.
  \eeq
  The last two terms on r.h.s are nonfactorizable, and their contributions
  are included in the effective color number $N_c^{eff}$. To
  evaluate the decay amplitudes, it is useful to introduce the combination of Wilson
  coefficients
  \beq
  a_{2i}^{eff} &=&
  c_{2i}^{eff}+\frac{1}{(N_c^{eff})_{2i}}c_{2i-1}^{eff},\non
  a_{2i-1}^{eff} &=&
  c_{2i-1}^{eff}+\frac{1}{(N_c^{eff})_{2i-1}}c_{2i}^{eff},
  \eeq
 The values of $N_c^{eff}$ can be found in \cite{hyc1},
  that is:
  \beq
  N_c^{eff}(V-A) \equiv (N_c^{eff})_1 \approx (N_c^{eff})_2 \approx
  (N_c^{eff})_3 \approx (N_c^{eff})_4 \approx (N_c^{eff})_9 \approx
  (N_c^{eff})_{10},\non
 N_c^{eff}(V+A) \equiv (N_c^{eff})_5 \approx (N_c^{eff})_6 \approx
  (N_c^{eff})_7 \approx (N_c^{eff})_8,
  \eeq
  As shown in \cite{hyc1} that in general $N_c^{eff}(V-A) \neq N_c^{eff}(V+A)$.
  The satisfied choice is that $N_c^{eff}(V-A) < 3 < N_c^{eff}(V+A)$.
  And it is reasonable to take the value of $N_c^{eff}(V-A)=2,
  N_c^{eff}(V+A)=5$. From now on, we will drop the superscript "eff" through the paper for convenience.

 \section{$B \to PV$ Decays in Model III 2HDM}

 Based on the effective Hamiltonian obtained via the operator product
 expansion and renormalization group evaluation, one can write
 down the amplitude for $B \to PV$ decays and calculate the branching
 ratios and CP violating asymmetries once a
 method is derived for computing the hadronic matrix elements. For
 purpose of this paper, we are going to explore the new physics contributions to the exclusive
 decays $B \to PV$ in the general model III 2HDM with spontaneous CP violation. For a numerical estimation,
 we will employ the generalized factorization approach described in the previous section.

 We begin with the following definitions for the branching ratio and CP violation asymmetry:
 \beq
 A_{CP} &=& \frac{|\bar{A}|^2-|A|^2}{|\bar{A}|^2+|A|^2},\non
 BR(B \to PV) &=&\frac{1}{2} \frac{p_c^3}{8\pi m_V^2}\tau_B(|\bar{A}|^2+|A|^2)/(\epsilon \cdot p_B)^2,
 \eeq
where $A$ and $\bar{A}$ are the decay amplitudes of $B$ and
$\bar{B}$ respectively, $\epsilon$ is the polarization vector of
the vector meson.
The input parameters in calculation are listed in Table.I.

\begin{table}[t]
\begin{center}
\caption{Input parameters} \label{para}
\vspace{0.2cm}
\begin{tabular}{|c|c|c|c|c|} \hline
$\tau_{B_d}$ & $\tau_{B_s}$&$M_{B_d}$ &$M_{B_s}$&$m_b$ \\
\hline 
$1.528\times10^{-12}ps$&$1.472\times 10^{-12}ps$ &
$5.28GeV$&$5.37GeV$&$4.2GeV$\\
\hline 
$m_t$&$m_u$&$m_d$&$m_c$&$m_s$\\
\hline 
$174GeV$&$3.2MeV$&$6.4MeV$&$1.1GeV$&$0.105GeV$\\
\hline 
$m_{\pi^{\pm}}$&$m_{\pi^0}$&$m_{\eta}$&$m_{\eta'}$&$m_{\rho^0}$\\
\hline 
$0.14GeV$&$0.135GeV$&$0.547GeV$&$0.958GeV$&$0.77GeV$\\
\hline 
$m_{\rho^{\pm}}$&$m_{\omega}$&$m_{\phi}$&$m_{K^{\pm}}$&$m_{K^0}$\\
\hline
$0.77GeV$&$0.782GeV$&$1.02GeV$&$0.494GeV$&$0.498GeV$\\
\hline 
$m_{K^{*\pm}}$&$m_{K^{*0}}$&$\Lambda_{QCD}$&$f_{\pi}$&$f_K$\\
\hline 
$0.892GeV$&$0.896GeV$&$225MeV$&$0.132GeV$&$0.16GeV$\\
\hline 
$f_{\rho}$&$f_{\omega}$&$f_{K^*}$&$f_{\phi}$&$f_{\rho}^T$\\
\hline
$0.21GeV$&$0.195GeV$&$0.221GeV$&$0.237GeV$&$0.147GeV$\\
\hline
$f_{\omega}^T$&$f_{K^*}^T$&$f_{\phi}^T$& & \\
\hline 
$0.133GeV$&$0.156GeV$&$0.183GeV$& &\\
\hline
\end{tabular}
\end{center}
\end{table}
 Here $f_M$ and $f_M^T$ are all decay constants of the mesons, $f_M$ comes from the
 experimental measurements, but $f_M^T$ is calculated from quenched
 lattice QCD and QCD sum rules\cite{ball3,lattice}.As for the form
 factor of pseudoscalar and vector mesons, we use the results from
 Light Cone Sum Rules(LCSR)\cite{ball1,ball2}, but for the form factor
 of $\eta'$, we use the value of BSW  model\cite{bsw}. And for the
 $\eta-\eta'$ mixing effects, we use the results of \cite{thuang}.
 The $B \to P(V)$ form factors' values are listed in
 Table.II. For comparison, we list both the results for light-cone sum
 rules(LCSR) and from sum rules in the framework of heavy quark
 effective field theory\cite{ybzuo}.

\begin{table}[t]
\begin{center}
\caption{The relevant Form Factors at $q^2 = 0$ for $B \to P$
transitions from LCSR\cite{ball1,ball2,ball3}(the first row),sum
rule in Heavy Quark Effective Field Theory\cite{ybzuo}(the second
row)and BSW model\cite{bsw}(the third row).The values in the
square brackets are the $B \to \eta'$ form factors.} \label{fac1}
\vspace{0.2cm}
\begin{tabular}{|c|c|c|c|c|c|c|}
\hline
\hline
 &Decay Channel&$B_q \to \pi$&$B_q \to K$&$B_q \to \eta^{(')}$&$B_s \to K$&$B_s \to \eta^{(')}$\\
\hline
&LCSR&$0.258$&$0.331$&$0.275[-]$&$-$&$-$\\
\cline{2-7}$F_0$&SRHQEFT&$0.285$&$0.345$&$0.247[-]$&$0.296$&$0.281[-]$\\
\cline{2-7}&BSW&$0.333$&$0.379$&$0.307[0.254]$&$0.274$&$0.335[0.282]$\\
\hline
\hline
\end{tabular}
\end{center}
\end{table}

\begin{table}[h]
\begin{center}
\caption{The relevant Form Factors at $q^2 = 0$ for $B \to V$
transitions from LCSR\cite{ball1,ball2,ball3}(the first row),sum
rule in Heavy Quark Effective Field Theory\cite{ybzuo}(the second
row)and BSW model\cite{bsw}(the third row).} \label{fac2}
\vspace{0.2cm}
\begin{tabular}{|c|c|c|c|c|c|c|}
\hline
\hline
  &Decay Channel&$B_q \to \rho$&$B_q \to \omega$&$B_q \to K^*$&$B_s \to \phi$&$B_s \to K^*$\\
\hline
&LCSR&$0.303$&$0.281$&$0.374$&$0.474$&$0.363$\\
\cline{2-7}$A_0$&SRHQEFT &$0.363$&$0.341$&$0.400$&$0.397$&$0.337$\\
\cline{2-7} &BSW&$0.281$&$0.280$&$0.321$&$0.475$&$0.364$\\
\hline
\hline
\end{tabular}
\end{center}
\end{table}

In the model III 2HDM, $\lambda_{ij}(i,j = c,s,b,t),
m_{H^\pm},m_{h_0},m_{A_0},m_{H_0}$ are free parameters that should
be constrained from experiments. It was shown from
$B_{d,s}^0-\bar{B}_{d,s}^0$ mixing that the parameters
$|\lambda_{cc}|$ and $|\lambda_{ss}|$ can reach to be around
$100$\cite{hcs3}, and their phases are not constrained too much.
In our present considerations, we simply fix their phases to be
$\pi/4$ to see their effects. For $\lambda_{tt}$ and
$\lambda_{bb}$, the constraints come from the experimental results
of $B-\bar{B}$ mixing, $\Gamma (b\to s\gamma)$,$\Gamma (b \to
c\tau\bar{\nu}_{\tau})$, $\rho_0,R_b$ and the electric dipole
moments (EDMS) of the electron and
neutron\cite{ARS,hcs1,hcs2,BCK,iltan}. For a numerical
calculation, we are going to consider the following three typical
parameter spaces which are allowed by the present experiments:
 \beq
 Case\quad A:\quad  |\lambda_{tt}|&=& 0.15; \quad  |\lambda_{bb}| = 50,\non
 Case\quad B:\quad  |\lambda_{tt}|&=& 0.3; \quad  |\lambda_{bb}| = 30,\non
 Case\quad C:\quad  |\lambda_{tt}|&=& 0.03; \quad |\lambda_{bb}| = 100,\non
 \eeq
 and:
 \beq
 \theta_{tt}+\theta_{bb} = \pi/2,
 \eeq
 For the Higgs mass, the following values are assumed:
 \beq
 & & m_{A_0} \simeq 120GeV, \quad  m_{h_0} \simeq 115GeV, \nonumber  \\
 & & m_{H_0} \simeq 160GeV, \quad m_{H^\pm} \simeq 200GeV
 \eeq

All the numerical results are presented in Table V $\sim$ Table IX.

\section{Conclusions and Discussions}

As the charged Higgs mediated one loop FCNC effects to the $\Delta B
= 1$ charmless decays are mostly characterized through the Wilson
coefficient $C_g^{eff}$, which is included in the
$C_{(3,4,5,6,7,8)}^{eff}$, and there are no new operators beyond the
basic operators $Q_{1,2...10}$, their contributions to the Wilson
coefficients are given in Table.III. On the contrary, the neutral
Higgs mediated processes will bring in new operators
$Q_{(11,12,\cdots,16)}$ with the new Wilson coefficients
$C_{(11,12,\cdots,16)}$. They are nonzero when the neutral Higgs
couples to the second and third generation of quarks, and the
numerical results are presented in Table.IV. From the above
calculations, it is seen that in some decay channels, the new
physics contributions can be significant, especially to CP
violations.

 a). As we have set the Yukawa couplings $\lambda_{iu}$ and $\lambda_{id}$
 to be zero, so the neutral Higgs contributions to
 $B \to (\rho, \omega, K^*)\pi, (\rho, \omega)K$ decays are actually ignored,
 only the charged Higgs give new contributions. One can see that
the branching ratio of $B \to \bar{K}^0\rho^0$ decay in the model
III 2HDM is the same as SM prediction (about $1.55\times 10^{-6}$),
which is far below the large central value of experimental result
$(5.4\pm0.9) \times 10^{-6}$. Though the annihilation diagram and
exchange diagram are not taken into account, their contributions are
still not enough to give such an enhancement. So one needs to find
some new mechanism to explain this discrepancy. The same situation
also appears in the $B \to K^-\rho^+, K^{*-}\pi^+$ decays, where the
experimental results are much larger than theoretical predictions
both in SM and 2HDM when simply using the generalized factorization
approach. Though the branching ratios could be enhanced by using
improved QCD Factorization(QCDF)\cite{qcdf1}, the resulting values
are still smaller than the measured results. b). The model III 2HDM
prediction for the CP violation of $B_d \to K\phi$ decay is $5\sim
7$ times larger than the SM prediction, which can be a signal to
look for new physics in future experiments. But the prediction of
the branching ratio are both smaller than the experimental one. c).
The SM and model III 2HDM predictions for branching ratio of $B_d
\to K^{*0}\pi^0$ are the same in size and all consistent with the
experimental result at $1\sigma$ level. While the new physics
prediction for CPV can flip the sign of the SM one and be $1\sim 5$
times larger in size and still within $1\sigma$ error of the
experiments. d). In $B_d \to K^{\ast}(\eta,\eta')$ decays, new
physics effects to CPV becomes significant. In $B \to K^{\ast}
\eta$, the 2HDM prediction is negative but the SM one is positive.
In $B \to K^{\ast}\eta'$, the 2HDM prediction can be as large as
$40\%$, which is about seven times of SM prediction. e). In $B_d \to
\rho^+\pi^-$ decay, the model III 2HDM prediction can enhance the CP
violation from about $-20\%$ in SM to about $-30\%$. Both SM and
2HDM predictions for the branching ratio of $B_d \to \rho^0\pi^0$
decay are much smaller than the experimental result.
 Such an inconsistence cannot be improved even in QCD factorization method\cite{qcdf1}.
 As for CP violation, the SM and model III 2HDM predictions have opposite sign with the magnitude
 $(10\sim15)\%$. As the current experimental error is still too big to draw a conclusion,
 much more precise measurement is needed to test it. f).  In $B_d \to \omega\pi$ decays,
 the new physics effects to CP violation may be distinct with the SM prediction,
 as it not only flips the sign but also enhances the magnitude
 by a factor of three.

 For $B_s^0 \to PV$ decays, the new physics contribution can be large in some decay channels. a). In $B_s \to K^*\eta$, the 2HDM
 prediction enhanced the direct CP violation to about $-50\%$ compared to the SM prediction $-28.8\%$,
 but for $B_s \to K^*\eta'$, the new physics contribution is destructive and reduce the SM
 prediction $-37\%$ to about $-20\%$. b). In $B_s \to \rho\eta^{(')}$
 decays, new physics contribution to branching ratio is
 destructive but gives an enhancement to the CPV to about four times of the SM prediction.
 c). In $B_s \to \phi\eta^{(')}$ decays, new physics effects to
 branching ratios and CPV are both significant. d). In $B_s \to
 K^0\bar{K}^{0\ast}$ decay, 2HDM can give about $25\sim70\%$ enhancement
 to the branching ratio.
 e). In $B \to K^0\phi$ decay, new physics effect to CPV is very significant, the SM prediction is almost zero,
 but the new physics effects can enhance it to about $-10\%$.

 For $B_u \to PV$ decays, there are also some new effects from the extra Higgs
 contributions:  a). In $B_u \to \pi^-\bar{K}^{*0}$ decay, the 2HDM prediction of CPV
 can be ten times of the SM one and are more closer to the
 experimental value. b). In $B_u \to K^-\phi$ decay, CPV can be about $10\%$ in 2HDM ,
 which is much larger than the SM prediction $1.44\%$
and is within $2\sigma$ level of experimental results.
 c). In $B_u \to K^{*-}\eta$ decays, new physics contribution reduced the CPV
 to about a half or a quarter of the SM one and is much closer to the experimental result.
d). In $B_u \to \rho^-\eta$ decay, 2HDM prediction for CPV is
$2\sim3$ times of the SM one and much closer to the experimental
 central value. e). In $B_u \to \pi^-\phi$ decay, new physics enhancement for the branching ratio and
 direct CPV can be all significant. f). In $B_u \to K^{*-}K^0$ decay, 2HDM predictions for CPV are $20\sim
 24\%$, which is very much larger than the SM prediction $-1.73\%$
On the contrary, in $B_u \to K^{*0}K^-$ decay, 2HDM prediction for
CPV can be much smaller than the SM one.

 From the above results, we see that in some decay channels, the
 theoretical predictions for branching ratios are still far from
 the experimental results not only in SM but also in model III 2HDM,such as
 $B \to K\rho,K^*\pi$ decays. And even using the improved QCDF, the situation
 cannot be improved much. There must be some new mechanism
 to improve those situations. For simplicity, we have not considered the possible effects of
 final state interaction (FSI) and the contributions from
 annihilation and exchange diagrams although they may play a
 significant rule in some decay channels. As for factorization part,
 in principle, $N_c^{eff}$ can vary from channel to channel as in
 the case of charm decay. However, in the energetic two-body B
 decays, $N_c^{eff}$ is expected to be process
 insensitive\cite{hyc1,hyc2}, and the preferred values are obtained
 from the data to be $N_c^{eff}(V-A) = 2, N_c^{eff}(V+A) = 5$\cite{hyc1,hyc2}.
 In a numerical calculation, we have considered only three cases for parameter
 choice in a general model III 2HDM to be consistent with the
 experimental results. Also we have totally neglected the first generation Yukawa couplings and
 the off-diagonal matrix elements of the Yukawa coupling matrix, such as
$\lambda_{tc, sb}$ \textit{etc}. to eliminate the FCNC at tree
level. However, it is still possible that FCNC involving the third
generation quarks exists at tree level, so the constraints can be
less stronger to get nonzero off-diagonal elements.

 In conclusion, we have shown that the new Higgs bosons in
 the general model III 2HDM with spontaneous CP violation can bring out some significant
 effects in some charmless B-meson decays, which can be good
 signals in the future B factory experiments to test the SM and look for new
 physics from more precise measurements.

\begin{acknowledgments}

This work was supported in part by the National Science Foundation
of China (NSFC) under the grant 10475105, 10491306, and the Project
of Knowledge Innovation Program (PKIP) of Chinese Academy of
Sciences.

\end{acknowledgments}


\newpage
\begin{appendix}
\section{The effective Wilson coefficients}
\begin{table}[h]
\begin{center}
\caption{The effective Wilson coefficients $C_{(1,2...10)}^{eff}$
in $b \to s$ process in SM and 2HDM at $\mu = m_b =4.2GeV$}
 \vspace{0.5cm}
\begin{tabular} {|c|c|c|c|c|}
\hline
Model & SM & Case A & Case B & Case C \\
\hline
$C_1^{eff}$&$1.17$&$1.17$&$1.17$&$1.17$\\
\hline
$C_2^{eff}$&$-0.37$&$-0.37$&$-0.37$&$-0.37$\\
\hline
$C_3^{eff}$&$0.024+0.0035I$&$0.024+0.006I$&$0.024+0.0048I$&$0.024+0.0075I$\\
\hline
$C_4^{eff}$&$-0.050-0.010I$&$-0.05-0.018I$&$-0.05-0.014I$&$-0.05-0.023I$\\
\hline
$C_5^{eff}$&$0.015+0.0035I$&$0.015+0.006I$&$0.015+0.005I$&$0.015+0.0075I$\\
\hline
$C_6^{eff}$&$-0.064-0.010I$&$-0.064-0.018I$&$-0.064-0.014I$&$-0.064-0.023I$\\
\hline
$C_7^{eff}$&$-0.00028-0.00024I$&$-0.00035-0.00024I$&$-0.00035-0.00024I$&$-0.00035-0.00024I$\\
\hline
$C_8^{eff}$&$0.00055$&$0.00061$&$0.00061$&$0.0006$\\
\hline
$C_9^{eff}$&$-0.011-0.00024I$&$-0.011-0.00024I$&$-0.011-0.00024I$&$-0.011-0.00024I$\\
\hline
$C_{10}^{eff}$&$0.0038$&$0.0034$&$0.0034$&$0.0034$\\
\hline
\end{tabular}\end{center}
\end{table}

 \begin{table}[h]
\begin{center}
\caption{The  Wilson coefficients $C_{(11,12...16)}^{eff}$ at $\mu
= m_b = 4.2GeV$}
 \vspace{0.5cm}
\begin{tabular} {|c|c|c|c|}
\hline
Parameter Space& Case A & Case B & Case C \\
\hline
$C_{11}^c$&$-0.089+0.12I$&$-0.089+0.19I$&$-0.11+0.13I$\\
\hline
$C_{12}^c$&$0$&$0$&$0$\\
\hline
$C_{13}^c$&$-0.031-0.051I$&$-0.054-0.072I$&$-0.030-0.055I$\\
\hline
$C_{14}^c$&$-0.00063-0.0010I$&$-0.0011-0.0015I$&$-0.000061-0.0011I$\\
\hline
$C_{15}^c$&$0.00035+0.00057I$&$0.00061+0.00080I$&$0.00034+0.00062I$\\
\hline
$C_{16}^c$&$-0.0011-0.00175I$&$-0.0019-0.0025I$&$-0.0010-0.0019I$\\
\hline
$C_{11}^s$&$-0.0085+0.012I$&$-0.0085+0.018I$&$-0.010+0.012I$\\
\hline
$C_{12}^s$&$0$&$0$&$0$\\
\hline
$C_{13}^s$&$-0.0030-0.0049I$&$-0.0052-0.0069I$&$-0.0029-0.0052I$\\
\hline
$C_{14}^s$&$-0.000060-0.00010I$&$-0.00011-0.00014I$&$-0.000059-0.00010I$\\
\hline
$C_{15}^s$&$0.000033+0.000055I$&$0.000058+0.000078I$&$0.000032+0.000059I$\\
\hline
$C_{16}^s$&$-0.00010-0.00017I$&$-0.00018-0.00024I$&$-0.0001-0.00018I$\\
\hline
\end{tabular}\end{center}
\end{table}

\newpage
\section{Numerical Results of $B \to PV$ Decays}
 \begin{table}[h]
\begin{center}
\caption{CP averaged branching ratios (in units of $10^{-6}$)(first
line) and direct CPV (second line)for charmless $B_d^0 \to PV$
decays in SM and 2HDM. $N_c^{eff}(V-A), N_c^{eff}(V-A)$ are fixed to
be $2$ and $5$ respectively and $N_c' = 3$. The parameter spaces
are: Case A: $(|\lambda_{tt}| = 0.15, |\lambda_{bb}| = 50, \theta =
\pi/2)$, Case B: $(|\lambda_{tt}| = 0.03, |\lambda_{bb}| = 100,
\theta = \pi/2)$, Case C: $(|\lambda_{tt}| = 0.3, |\lambda_{bb}| =
30, \theta = \pi/2)$. And $\lambda_{cc} = \lambda_{ss} =
100e^{i\pi/4}$. }
 \vspace{0.5cm}
\begin{tabular} {|c|c|c|c|c|c|}
\hline
Decay channel & SM & Case A (2HDM) & Case B (2HDM)& Case C (2HDM) & Exp \\
\hline
$B_d^0 \to K^0\rho^0$&$1.56$&$1.55$&$1.55$&$1.56$&$5.4\pm0.9$\\
\hline
$ $&$1.7\%$&$2.10\%$&$1.97\%$&$2.20\%$&$-$\\
\hline
$B_d^0 \to K^-\rho^+$&$2.01$&$1.94$&$1.97$&$1.91$&$9.9_{-1.5}^{+1.6}$\\
\hline
$ $&$-3.6\%$&$-3.83\%$&$-3.90\%$&$-3.76\%$&$(17^{+15}_{-16})\%$\\
\hline
$B_d^0 \to K^{*-}\pi^+$&$3.24$&$3.92$&$3.56$&$4.33$&$9.8\pm1.1$\\
\hline
$ $&$24.5\%$&$27.5\%$&$26.2\%$&$28.2\%$&$(-5\pm14)\%$\\
\hline
$B_d^0 \to K^{*0}\pi^0$&$1.47$&$1.47$&$1.46$&$1.51$&$1.7\pm0.8$\\
\hline
$ $&$-2.50\%$&$7.20\%$&$2.30\%$&$11.5\%$&$(-1^{+27}_{-26})\%$\\
\hline
$B_d^0 \to K^0\phi$&$4.80$&$5.22$&$5.23$&$5.18$&$8.3_{-1.0}^{+1.2}$\\
\hline
$ $&$1.40\%$&$5.96\%$&$10.0\%$&$10.3\%$&$-$\\
\hline
$B_d^0 \to K^*\eta$&$9.41$&$10.4$&$10.7$&$10.8$&$16.1\pm1.0$\\
\hline
$ $&$ 1.86\%$&$-2.68\%$&$-3.85\%$&$-1.93\%$&$(19\pm5)\%$\\
\hline
$B_d^0 \to K^*\eta'$&$1.33$&$1.18$&$1.49$&$1.20$&$3.8\pm1.2$\\
\hline
$ $&$ 5.50\%$&$32.7\%$&$22.1\%$&$40.4\%$&$(-8\pm25)\%$\\
\hline
$B_d^0 \to K^0\omega$&$0.43$&$0.44$&$0.44$&$0.44$&$4.8\pm0.6$\\
\hline
$ $&$0.00$&$0.33\%$&$0.32\%$&$0.32\%$&$-$\\
\hline
$B_d^0 \to \rho^-\pi^+$&$15.8 $&$15.3 $&$15.1 $&$15.1 $&$24.0\pm2.5 $\\
\hline
$ $&$ -4.3\%$&$-4.4\%$&$-4.3\%$&$-4.4\%$&$-$\\
\hline
\end{tabular}\end{center}
\end{table}
\newpage

\begin{table}[h]
\begin{center}
\caption{Continue Table III }
 \vspace{0.5cm}
\begin{tabular} {|c|c|c|c|c|c|}
\hline
Decay channel & SM & Case A (2HDM) & Case B (2HDM)& Case C (2HDM) & Exp \\
\hline
$B_d^0 \to \rho^+\pi^-$&$17.3$&$16.3$&$17.7$&$15.0$&$24.0\pm2.5 $\\
\hline
$ $&$ -18.8\%$&$-26.5\%$&$-26.4\%$&$-26.4\%$&$-$\\
\hline
$B_d^0 \to K^{0*}\bar{K^0}$&$0.23$&$0.24$&$0.24$&$0.25$&$<1.9$\\
\hline
$ $&$-11.5\%$&$-10.1\%$&$-14.6\%$&$-6.45\%$&$-$\\
\hline
$B_d^0 \to K^{0}\bar{K^{*0}}$&$0.038$&$0.037$&$0.036$&$0.037$&$-$\\
\hline
$ $&$ -1.73\%$&$20.3\%$&$22.0\%$&$24.3\%$&$-$\\
\hline
$B_d^0 \to \phi\eta$&$0.0039$&$0.0038$&$0.0038$&$0.0038$&$<0.6$\\
\hline
$ $&$1.13\%$&$1.35\%$&$1.25\%$&$1.45\%$&$ $\\
\hline
$B_d^0 \to \phi\eta'$&$0.0023$&$0.0022$&$0.0024$&$0.0022$&$<1.0$\\
\hline
$ $&$1.13\%$&$1.35\%$&$1.25\%$&$1.45\%$&$-$\\
\hline
$B_d^0 \to \phi\pi$&$0.015$&$0.0017$&$0.0017$&$0.0016$&$<0.28$\\
\hline
$ $&$3.90\%$&$1.35\%$&$1.25\%$&$1.45\%$&$-$\\
\hline
$B_d^0 \to \rho^0\pi^0$&$0.80$&$0.81$&$0.82$&$0.80$&$1.8^{+0.6}_{-0.5}$\\
\hline
$ $&$-10.5\%$&$14.4\%$&$13.7\%$&$14.9\%$&$(-49^{+70}_{-83})\%$\\
\hline
$B_d^0 \to \rho\eta$&$0.82$&$0.88$&$0.83$&$0.92$&$<1.5$\\
\hline
$ $&$12.3\%$&$6.93\%$&$3.21\%$&$6.91\%$&$-$\\
\hline
$B_d^0 \to \rho\eta'$&$0.50$&$0.55$&$0.54$&$0.57$&$<3.7$\\
\hline
$ $&$5.88\%$&$6.33\%$&$6.87\%$&$6.20\%$&$-$\\
\hline
$B_d^0 \to \omega\pi$&$0.60$&$0.56$&$0.53$&$0.59$&$<1.2$\\
\hline
$ $&$-4.97\%$&$12.4\%$&$12.7\%$&$12.1\%$&$-$\\
\hline
$B_d^0 \to \omega\eta$&$0.84$&$0.77$&$0.80$&$0.73$&$<1.9$\\
\hline
$ $&$-13.9\%$&$-11.1\%$&$-9.04\%$&$-11.2\%$&$-$\\
\hline
$B_d^0 \to \omega\eta'$&$0.53$&$0.46$&$0.50$&$0.44$&$<2.8$\\
\hline
$ $&$-19.7\%$&$-25.0\%$&$-26.0\%$&$-26.1\%$&$-$\\
\hline
\end{tabular}\end{center}
\end{table}

\newpage

\begin{table}[h]
\begin{center}
\caption{CP averaged branching ratios (in units of
$10^{-6}$)(first line) and direct CPV (second line)for charmless
$B_s^0 \to PV$ decays in SM and 2HDM. }
 \vspace{0.5cm}
\begin{tabular} {|c|c|c|c|c|}
\hline
Decay channel & SM & Case A (2HDM)& Case B (2HDM)& Case C (2HDM)  \\
\hline
$B_s^0 \to K^{*+}\pi^-$&$8.34$&$8.73$&$8.34$&$8.44$\\
\hline
$ $&$-0.13\%$&$-0.12\%$&$-0.13\%$&$-0.13\%$\\
\hline
$B_s^0 \to K^+\rho^-$&$26.7$&$27.7$&$26.5$&$26.2$\\
\hline
$ $&$-4.3\%$&$-4.3\%$&$-4.3\%$&$-4.4\%$\\
\hline
$B_s^0 \to K^{*0}\pi^0$&$0.21$&$0.20$&$0.21$&$0.21$\\
\hline
$ $&$5.0\%$&$5.1\%$&$5.1\%$&$5.1\%$\\
\hline
$B_s^0 \to \rho K^0$&$0.59$&$0.70$&$0.66$&$0.73$\\
\hline
$ $&$16.5\%$&$14.2\%$&$14.7\%$&$13.6\%$\\
\hline
$B_s^0 \to K^0\omega$&$0.70$&$0.73$&$0.69$&$0.68$\\
\hline
$ $&$-18.3\%$&$-18.3\%$&$-18.4\%$&$-18.1\%$\\
\hline
$B_s^0 \to K^*\eta$&$0.29$&$0.32$&$0.31$&$0.33$\\
\hline
$ $&$-28.8\%$&$-43.2\%$&$-49.9\%$&$-42.8\%$\\
\hline
$B_s^0 \to K^*\eta'$&$0.20$&$0.23$&$0.21$&$0.25$\\
\hline
$ $&$-37.1\%$&$-21.2\%$&$-21.3\%$&$-17.7\%$\\
\hline
$B_s^0 \to K^-K^{*+}$&$1.98$&$2.27$&$2.24$&$2.31$\\
\hline
$ $&$-3.6\%$&$-3.3\%$&$-3.4\%$&$-3.2\%$\\
\hline
$B_s^0 \to K^+K^{*-}$&$5.49$&$6.97$&$6.81$&$7.02$\\
\hline
$ $&$24.5\%$&$22.8\%$&$20.3\%$&$21.6\%$\\
\hline
$B_s^0 \to \rho\eta$&$0.21$&$0.04$&$0.04$&$0.04$\\
\hline
$ $&$3.6\%$&$18.4\%$&$18.4\%$&$18.3\%$\\
\hline
$B_s^0 \to \rho\eta'$&$0.12$&$0.03$&$0.02$&$0.03$\\
\hline
$ $&$3.6\%$&$18.3\%$&$18.4\%$&$18.4\%$\\
\hline
$B_s^0 \to \phi\pi$&$0.21$&$0.04$&$0.03$&$0.04$\\
\hline
$ $&$0.00\%$&$0.00$&$0.00$&$0.00$\\
\hline
$B_s^0 \to \omega\eta$&$0.02$&$0.02$&$0.02$&$0.02$\\
\hline
$ $&$43.1\%$&$40.0\%$&$39.6\%$&$40.1\%$\\
\hline
$B_s^0 \to \omega\eta'$&$0.01$&$0.01$&$0.01$&$0.01$\\
\hline
$ $&$43.1\%$&$40.0\%$&$39.6\%$&$40.1\%$\\
\hline
$B_s^0 \to \phi\eta$&$12.2$&$17.9$&$17.7$&$20.0$\\
\hline
$ $&$2.2\%$&$-12.3\%$&$-10.6\%$&$-14.2\%$\\
\hline
$B_s^0 \to \phi\eta'$&$0.61$&$1.68$&$1.92$&$2.23$\\
\hline
$ $&$15.6\%$&$-21.2\%$&$-21.3\%$&$-17.7\%$\\
\hline
$B_s^0 \to K^0\bar{K^*}$&$5.78$&$8.03$&$7.00$&$9.14$\\
\hline
$ $&$1.27\%$&$4.10\%$&$2.29\%$&$5.0\%$\\
\hline
$B_s^0 \to  K^{*0}\bar{K}$&$1.28$&$1.03$&$1.00$&$1.06$\\
\hline
$ $&$1.0\%$&$1.2\%$&$2.2\%$&$1.0\%$\\
\hline
$B_s^0 \to \phi K^0$&$0.14$&$0.12$&$0.12$&$0.12$\\
\hline
$ $&$0.23\%$&$-9.7\%$&$-8.8\%$&$-11.7\%$\\
\hline
\end{tabular}\end{center}
\end{table}

\newpage

\begin{table}[t]
\begin{center}
\caption{CP averaged branching ratios (in units of
$10^{-6}$)(first line) and direct CPV (second line)for charmless
$B_u^- \to PV$ decays in SM and 2HDM. }
 \vspace{0.5cm}
\begin{tabular} {|c|c|c|c|c|c|}
\hline
Decay channel & SM & Case A 2HDM) & Case B (2HDM)& Case C (2HDM) & Exp.\\
\hline
$B_u^- \to K^-\rho^0$&$0.59$&$0.58$&$0.58$&$0.56$&$4.25^{+0.55}_{-0.56}$\\
\hline
$ $&$-1.88\%$&$-2.40\%$&$-2.32\%$&$-2.38\%$&$(31^{+11}_{-10})\%$\\
\hline
$B_u^- \to K^{*-}\pi^0$&$2.62$&$2.80$&$3.25$&$2.80$&$6.9\pm2.3$\\
\hline
$ $&$18.9\%$&$22.3\%$&$20.4\%$&$22.3\%$&$(4\pm29)\%$\\
\hline
$B_u^- \to \bar{K^0}\rho^-$&$1.09$&$1.10$&$1.10$&$1.10$&$<48$\\
\hline
$ $&$0.34\%$&$1.15\%$&$0.75\%$&$1.15\%$&$-$\\
\hline
$B_u^- \to \pi^-\bar{K^*}^0$&$3.67$&$3.88$&$3.66$&$3.88$&$11.3\pm1.0$\\
\hline
$ $&$-1.28\%$&$-12.9\%$&$-5.41\%$&$-12.9\%$&$(-8.6\pm5.6)\%$\\
\hline
$B_u^- \to K^-\omega$&$2.22$&$2.17$&$2.20$&$2.17$&$6.9\pm0.5$\\
\hline
$ $&$0.00\%$&$-0.33\%$&$-0.32\%$&$-0.33\%$&$(5\pm6)\%$\\
\hline
$B_u^- \to K^-\phi$&$5.16$&$5.92$&$5.57$&$5.93$&$8.30\pm{0.65}$\\
\hline
$ $&$1.44\%$&$11.7\%$&$10.0\%$&$10.3\%$&$(3.4\pm4.4)\%$\\
\hline
$B_u^- \to K^{*-}\eta$&$9.36$&$10.51$&$11.09$&$10.6$&$19.5^{+1.6}_{-1.5}$\\
\hline
$ $&$13.6\%$&$6.73\%$&$10.2\%$&$4.52\%$&$(2\pm6)\%$\\
\hline
$B_u^- \to K^{*-}\eta'$&$1.53$&$1.32$&$1.38$&$1.37$&$4.9^{+2.1}_{-1.9}$\\
\hline
$ $&$52.2\%$&$55.4\%$&$52.4\%$&$56.9\%$&$(30^{+33}_{-37})\%$\\
\hline
$B_u^- \to \pi^0\rho^-$&$11.4$&$11.1$&$11.3$&$11.1$&$10.8^{+1.4}_{-1.5}$\\
\hline
$ $&$-3.0\%$&$-3.1\%$&$-3.0\%$&$-3.1\%$&$(2\pm11)\%$\\
\hline
$B_u^- \to \pi^-\rho^0$&$7.36$&$7.75$&$7.50$&$7.75$&$8.7^{+1.0}_{-1.1}$\\
\hline
$ $&$4.2\%$&$4.1\%$&$4.2\%$&$4.1\%$&$(-7^{+12}_{-13})\%$\\
\hline
$B_u^- \to \pi^-\omega$&$6.85$&$6.50$&$6.65$&$6.50$&$6.7\pm0.6$\\
\hline
$ $&$-4.7\%$&$-4.8\%$&$-4.7\%$&$-4.8\%$&$(-4\pm7)\%$\\
\hline
$B_u^- \to \rho^-\eta$&$11.1$&$13.2$&$11.0$&$10.8$&$5.3^{+1.2}_{-1.1}$\\
\hline
$ $&$-0.90\%$&$-2.35\%$&$-3.34\%$&$-2.36\%$&$(1\pm16)\%$\\
\hline
$B_u^- \to \rho^-\eta'$&$14.0$&$12.7$&$13.7$&$12.8$&$9.1^{+3.7}_{-2.8}$\\
\hline
$ $&$-9.9\%$&$-10.1\%$&$-9.60\%$&$-10.1\%$&$(-4\pm28)\%$\\
\hline
$B_u^- \to \pi^-\phi$&$0.0036$&$0.016$&$0.016$&$0.016$&$<0.24$\\
\hline
$ $&$1.13\%$&$15.5\%$&$7.96\%$&$15.5\%$&$-$\\
\hline
$B_u^- \to K^{*-}K^0$&$0.038$&$0.037$&$0.036$&$0.037$&$-$\\
\hline
$ $&$-1.73\%$&$20.3\%$&$22.0\%$&$24.3\%$&$-$\\
\hline
$B_u^- \to K^-K^{*0}$&$0.25$&$0.27$&$0.26$&$0.27$&$<5.3$\\
\hline
$ $&$-37.1\%$&$-5.13\%$&$-14.6\%$&$-6.45\%$&$-$\\
\hline
\end{tabular}\end{center}
\end{table}
\end{appendix}

\begin{thebibliography}{35}
\bibitem{KM}M.Kobayashi and T.Maskawa, Prog.Theor.Phys.{\bf 49},652(1973).

\bibitem{YLWU} Y.L.Wu, Phys.Rev.{\bf D64},016001(2001),
hep-ph/0012371; \\
For a review also see: S. Bertolini, "Theory
Status of $\varepsilon'/\varepsilon$ ", Frascati Phys. Ser. 28
275-290 ги2002), hep-ph/0206095.

\bibitem{KK} Alavi-Harati, et al. [KTeV Collaboration], Phys. Rev. {\bf D67} 012005,
(2003); \\  J.R. Batley, et al. [NA48 Collaboration], Phys. Lett.
{\bf B544} 97 (2002).

\bibitem{WZZ} Y.L. Wu and Y.F. Zhou, Phys.Rev. {\bf D71} 021701 (2005),
hep-ph/0409221; \\
Y.L. Wu, Y.F. Zhou and C. Zhuang, Phys.Rev. {\bf D74} 094007
(2006), hep-ph/0609006.

\bibitem{BB}
K.~Abe {\it et al.}  [Belle Collaboration], Phys. Rev. Lett. {\bf
93}, 021601 (2004); \\
 B.~Aubert {\it et al.}  [BaBar
Collaboration], Phys. Rev. Lett. {\bf 93}, 131801 (2004).

\bibitem{TDL} T.D. Lee, Phys. Rev. {\bf D8}, 1226 (1973); Phys. Rep. {\bf 9},
143 (1974).

\bibitem{HS}
The Higgs Hunter's Guide by J.Gunion et al., Addison Wesley, New
York, 1990.
P. Sikivie, Phys. Lett. {\bf B65}, 141 (1976); \\
S.L.Glashow and S.Weinberg, Phys. Rev. {\bf D15}, 1958(1977);\\
H.E. Haber, G.L. Kane and T. Sterling, Nucl. Phys. {\bf B161}, 493 (1979); \\
N.G. Deshpande and E. Ma, Phys. rev. {\bf D18}, 2574 (1978); \\
H. Georgi, Hadronic J. {\bf 1}, 155 (1978); \\
J.F. Donoghue and L.-F. Li, Phys. Rev. {\bf D19}, 945 (1979); \\
A.B. Lahanas and C.E. Vayonakis, Phys. Rev. {\bf D19}, 2158 (1979); \\
L.F. Abbott, P. Sikivie and M.B. Wise, Phys. Rev. {\bf D21},
1393 (1980); \\
G.C. Branco, A.J. Buras and J.M. Gerard, Nucl. Phys. {\bf B259},
306 (1985); \\
B. McWilliams and L.-F. Li, Nucl. Phys. {\bf B179}, 62 (1981); \\
J.F. Gunion and H.E. Haber, Nucl. Phys. {\bf B272}, 1 (1986); \\
J. Liu and L. Wolfenstein, Nucl. Phys. {\bf B289}, 1 (1987);\\
J.F.Cheng,C.S.Huang and X.H.Wu, Nucl. Phys. {\bf B701},54,(2004);\\
R.Barbieri and L.J.Hall,hep-ph/0510243.
\bibitem{TH1}
T.P.Cheng and M.Sher,Phys. Rev. {\bf D 35},3484(1987); M.Sher and
Y.Yao,Phys. Rev. {\bf D44}, 1461(1991);
\bibitem{TH2} A. Antaramian, L.J. Hall, and A. Rasin, Phys. Rev. Lett. {\bf 69}, 1871 (1992).
\bibitem{TH3} W.S. Hou, Phys. Lett. {\bf B296}, 179 (1992);  \\
D.W. Chang, W.S. Hou and W.Y. Keung, Phys. Rev. {\bf D48}, 217 (1993); \\
M.J. Savage, Phys. Lett. {\bf B266}, 135 (1991).
\bibitem{TH4} L.J. Hall and S. Weinberg, Phys. Rev. {\bf D48}, 979 (1993).
\bibitem{YLW1} Carnegie-Mellon Univ. report, CMU-HEP94-01, hep-ph/9404241, 1994 (unpublished); \\
Y.L. Wu,  in: Proceedings at 5th Conference on the Intersections of
Particle and Nuclear Physics, St. Petersburg, FL, 31 May- 6 Jun
1994, pp338, edited by S.J. Seestrom (AIP, New York, 1995),
hep-ph/9406306.
\bibitem{YLW2}
Y.L.Wu and L.Wolfenstein, Phys.Rev.Lett 73,1762(1994).
\bibitem{ARS}D.Atwood, L.Reina and A. Soni, phys. Rev. {\bf D55}, 3156 (1997).
\bibitem{dai}
Y.B.Dai, C.S.Huang and H.W.Huang, Phys.Lett.{\bf B390},257 (1997).

\bibitem{BCK} D. Bowser-Chao , K. Cheung, and Wai-Yee Keung, Phys.Rev. {\bf D59} 115006, (1999).

\bibitem{KSW} K. Kiers, A. Soni, and G.H. Wu, Phys. Rev. {\bf D59},096001(1999).
%

\bibitem{WZ} Y.L. Wu and Y.F. Zhou, Phys.Rev. {\bf D61} 096001 (2000),
hep-ph/9906313.

\bibitem{WW} L. Wolfenstein and Y.L. Wu, Phys. Rev. Lett., {\bf 73}, 2809 (1994).
\bibitem{YLW3} Y.L. Wu, Chin. Phys. Lett. {\bf 16} 339 (1999),
 hep-ph/9805439.

\bibitem{hcs3}
C.S.Huang, J.T.Li, Int.J.Mod.Phys.A20,161(2005).
\bibitem{LO1} G. Buchalla, A.J. Buras and M.K. Harlander, Nucl. Phys. {\bf B337} (1990) 313
\bibitem{LO2} E.A. Paschos and Y.L. Wu, Mod. Phys. Lett. {\bf A6} (1991) 93.

\bibitem{buchalla1}
G.Buchalla, A.J.Buras and
M.E.Lautenbacher,Rev.Mod.Phys.68,(1125)1996;\\
A. Buras, M. Jamin, M. Lautenbacher and P.H. Weisz,
 Nucl. Phys. {\bf B400} (1993) 37, 75; \\
A. Buras, M. Jamin and M. Lautenbacher, Nucl. Phys. {\bf B408}
(1993) 209; \\
M. Cuichini, E. Franco, G. Martinelli and L. Reina,
 Phys. Lett. {\bf B301} (1993) 263;\\
 M. Cuichini et al.,  Nucl. Phys. {\bf B415} (1994) 403.

\bibitem{hcs1}
C.S.~Huang, and S.H.~Zhu, Phys. Rev. {\bf D68},114020(2003).

\bibitem{wise}
B. Grinstein, R. Springer and M.B. Wise, Nucl. Phys. {\bf B339},
269 (1990); A. Ali and C. Greub, Phys. Lett. {\bf B259}, 182
(1991).



\bibitem{xiao1}
Z.J.Xiao, C.S.Li, Phys. Rev. {\bf D63},074005(2001).

\bibitem{zhuang}
Z.J.Xiao, C.Zhuang, \epjc 33,349(2004).

\bibitem{hcs2}
Y.B.Dai,C.S.Huang, J.T.Li and W.J.Li, Phys. Rev. {\bf
D67},096007(2003).

\bibitem{hyc2}
H.Y.Cheng,B.Tseng, Phys. Rev. {\bf D58},094005(1998).

\bibitem{lucd}
A.Ali, G.Kramer and C.D.L\"{u}, Phys. Rev. {\bf D59},
014005,(1999).

\bibitem{hyc1}
Y.H.Chen, H.Y.Cheng, B.Tseng and K.C.Yang, Phys. Rev. {\bf
D60},094014(1999).

\bibitem{ali}
A.Ali, G.Kramer and C.D.L\"{u}, Phys. Rev. {\bf D58},
094009(1998).


\bibitem{du}
D.S.Du, Y.D.Yang and G.H.Zhu, Phys. Rev. {\bf D59},014007(1999).

\bibitem{leutwyler}
H.Leutwyler, Nucl.Phys {\bf B} (Proc.Suppl.) 64,223(1998).

\bibitem{ball2}
P.Ball and R.Zwicky, Phys. Rev. {\bf D71},014015(2005).

\bibitem{ball3}
P.Ball and M.Boglione, Phys. Rev. {\bf D68},094006(2003).

\bibitem{lattice}
D.Becirevic et al., JHEP 0305(2003) 007.


\bibitem{ball1}
P.Ball and R.Zwicky, Phys. Rev. {\bf D71},014029(2005).


\bibitem{bsw}
M. Wirbel, B.Stech and M.Bauer, Z, Phys. C29,637(1985). M.Bauer,
B.Stech and M. Wirbel, Z, Phys. C34,103(1987).



\bibitem{thuang}
J.Cao,F.G.Cao, T.Huang and B.Q.Ma, Phys.Rev.{\bf
D58},113006(1998).

\bibitem{ybzuo}
Y.L.Wu, M.Zhong, Y.B.Zuo,  Int. J. Mod. Phys. {\bf A21} 6125
(2006); hep-ph/0604007.

\bibitem{iltan}
E.O.Iltan, J.Phys.G27,1723,(2001).

\bibitem{qcdf1}
M.Beneck and M.Neubert, Nucl.Phys {\bf B675},333(2003).

\end{thebibliography}
\end{document}